\documentclass[12pt]{article}

\RequirePackage[OT1]{fontenc}
\usepackage{amsthm,amsmath,natbib}
\usepackage{array}
\usepackage{float}
\usepackage{booktabs}
\usepackage{tabularx}
\usepackage{xcolor}
\usepackage{color}
\RequirePackage[colorlinks,citecolor=blue,urlcolor=blue]{hyperref}
\usepackage{alltt}
\usepackage{fancyvrb}
\usepackage{gensymb}
\usepackage{algorithm2e}
\usepackage{pdfpages}
\usepackage{multirow}
\usepackage{lineno}

\def\bi{\begin{itemize}}
\def\ei{\end{itemize}}
\def\be{\begin{equation}}
\def\ee{\end{equation}}

\def\rm#1{\mathrm{#1}}

\usepackage[english]{babel}
\usepackage{amsfonts}
\usepackage{amssymb}
\usepackage{graphicx}
\usepackage{enumerate}
\usepackage{bm}

\def\iid{{\ {\buildrel \rm{iid}\over \sim}\ }}

\renewcommand{\vec}[1]{\boldsymbol{#1}}

\newcommand{\blind}{1}

\begin{document}
\thispagestyle{empty}
\baselineskip=27pt
\vskip 4mm
\begin{center} {\Large{\bf Geostatistical modeling to capture seismic-shaking patterns from earthquake-induced landslides}}
\end{center}

\baselineskip=12pt
\vskip 3mm

\if1\blind
{
\begin{center}
\large
Luigi Lombardo$^{1,2*}$, Haakon Bakka$^1$, Hakan Tanyas$^3$, Cees van Westen$^3$, P. Martin Mai$^2$, Raphael Huser$^1$
\end{center}

\footnotetext[1]{
\baselineskip=10pt King Abdullah University of Science and Technology (KAUST), Computer, Electrical and Mathematical Sciences and Engineering (CEMSE) Division, Thuwal 23955-6900, Saudi Arabia.}
\footnotetext[2]{
\baselineskip=10pt King Abdullah University of Science and Technology (KAUST), Physical Sciences and Engineering (PSE) Division, Thuwal 23955-6900, Saudi Arabia.}
\footnotetext[3]{
\baselineskip=10pt University Twente, Faculty of Geo-Information Science and Earth Observation (ITC), The Netherlands}
} \fi

\baselineskip=16pt
\vskip 2mm
\centerline{\today}
\vskip 4mm

\begin{center}
{\large{\bf Abstract}}
\end{center}

In this paper, we investigate earthquake-induced landslides using a geostatistical model that includes a latent spatial effect (LSE). The LSE represents the spatially structured residuals in the data, which are complementary to the information carried by the covariates. To determine whether the LSE can capture the residual signal from a given trigger, we test whether the LSE is able to capture the pattern of seismic shaking caused by an earthquake from the distribution of seismically induced landslides, without prior knowledge of the earthquake being included in the statistical model. We assess the landslide intensity, i.e., the expected number of landslide activations per mapping unit, for the area in which landslides triggered by the Wenchuan (M 7.9, May 12, 2008) and Lushan (M 6.6, April 20, 2013) earthquakes overlap. We chose an area of overlapping landslides in order to test our method on landslide inventories located in the near and far fields of the earthquake. We generated three different models for each earthquake-induced landslide scenario: \textit{i}) seismic parameters only (as a proxy for the trigger); \textit{ii}) the LSE only; and \textit{iii}) both seismic parameters and the LSE. The three configurations share the same set of morphometric covariates. This allowed us to study the pattern in the LSE and assess whether it adequately approximated the effects of seismic wave propagation. Moreover, it allowed us to check whether the LSE captured effects that are not explained by the shaking levels, such as topographic amplification. Our results show that the LSE reproduced the shaking patterns in space for both earthquakes with a level of spatial detail even greater than the seismic parameters. In addition, the models including the LSE perform better than conventional models featuring seismic parameters only.     
\baselineskip=16pt

\par\vfill\noindent

\newpage
\baselineskip=16pt

\section{Introduction}

Substantial improvements have been made in landslide predictive mapping during the last two decades \citep{Reichenbach2018}. Research has progressed from simple heuristic approaches \citep[e.g.,][]{leoni2009gis}, towards deterministic methods \citep{Bout2018}, multivariate statistics \citep[e.g.,][]{lombardo2014test}, and data mining techniques \citep[e.g.,][]{LEE2018}. However, the estimation target (i.e., the landslide susceptibility) has remained the same. As a community, we build models based on presence-absence data to predict \emph{where} future landslides may occur \citep{Guzzetti2006}.

In other words, landslide data are commonly managed in a binary framework, which is often modeled using the Bernoulli distribution \citep[e.g.,][]{Camilo2017}. Although this approach has proven to be useful and robust, some information is lost by restricting the paradigm to presence and absence. \citet{Robinson2017} and \citet{Lombardo.etal:2018} proposed shifting the statistical paradigm by modeling landslides by using their count per mapping unit, rather than their presence and absence. \citet{Robinson2017} proposed using this framework for the rapid assessment of coseismic landslide initiations with a Fuzzy Logic algorithm. Conversely, \citet{Lombardo.etal:2018} proposed a method for analyzing the number of landslide activations across a geographic space using a log-Gaussian Cox process, and explaining the data according to a statistical model based on the Poisson distribution. Thus, the resulting predictive map reflects the \emph{intensity} rather than the \emph{susceptibility}, jointly answering the questions \emph{where} and \emph{how many} landslides are likely to occur in a given region using the same model.

Since this approach has only been tested on landslides triggered by rainfall, the case of earthquake-induced landslides remains unexplored. Earthquakes are disastrous natural processes that occur in tectonically active areas \citep{isacks1968seismology}, and the damage of earthquakes combined with subsequent landslides may extend over large regions \citep{Fan2018}. The increasing density of seismic networks over past decades, generally with a good spatial coverage, has increased the quantity and quality of data available for estimating ground motion in areas that have experienced an earthquake \citep{trifunac2001}. Conversely, precipitation rates may vary over short distances even within a small watershed, and rain-gauge networks may not be available to measure rainfall discharges across time and space at fine resolution \citep{aronica2012flash}, although satellite rainfall measurements using new products like Global Precipitation Measurement (GPM) may improve this substantially \citep{Smith2007}. This is the main reason why susceptibility models that spatially predict seismically induced landslides include ground motion parameters, such as peak ground acceleration (PGA) or peak ground velocity (PGV), among the covariate set \citep[see, e.g.,][]{UMAR2014,Parker2015}; whereas in the case of storm-induced mass movements, the precipitation amount is hardly ever considered \citep{cama2017improving,lombardo2016b}. 

\citet{Lombardo.etal:2018} presented an innovative approach to account for missing covariates in landslide prediction studies. In their work, when information about the trigger is missing, they hypothesized that the spatially-coherent residual component in a model can be used to reconstruct the spatial pattern of the triggering rainfall. The spatial residuals represent the spatial structure of the data not modeled by the common covariates, and they can be captured via a latent spatial effect (LSE). 

This hypothesis is justified because the storm signal should dominate the landslide distribution over space compared to the geomorphological factors. However, due to the lack of raingauges in their study area, this assumption could not be rigorously verified. Therefore, here we test the approach of \citet{Lombardo.etal:2018} on two earthquake-triggered landslide inventories for which shaking-level data are available.

In particular, we selected the area where the landslide inventories caused by the Wenchuan (M 7.9, May 12, 2008) and Lushan (M 6.6, April 20, 2013) earthquakes overlap. Then we used the subsets of the original inventories falling within area to build separate landslide intensity models, with the goal of predicting the number of landslides per mapping unit under analogous triggering conditions \citep{Lombardo.etal:2018}. For each dataset, we generated models that alternatively included seismic parameters or the LSE, while keeping the morphometric and thematic covariates constant, to check whether the LSE is capable of adequately approximating the shaking-level patterns. In addition, we ran a third model, in which both seismic parameters and the LSE were included, to determine whether any residual effects over space remained when the ground motion was already accounted for in the model. By testing two scenarios, both the Lushan and the Wenchuan earthquakes, we aimed to verify that the LSE is capable of approximating shaking-level patterns respectively in the near and far fields of a seismogenic source.

The remainder of this paper is organized as follows.~In \S\ref{sec:data}, we briefly describe the Lushan and Wenchuan disasters and how we constructed the dataset. In \S\ref{sec:popland}, we explain how we built the two sets of landslide intensity models. In \S\ref{sec:results}, we present the results. In \S\ref{sec:Discussion}, we interpret and discuss the results. In \S\ref{sec:Conclusions}, we add concluding remarks and suggest further improvements and future challenges. 

\section{Dataset Creation}
\label{sec:data}
\subsection{Wenchuan and Lushan Inventories}
\label{sec:data.description}

\begin{figure}[t!]
	\centering
	\includegraphics[width=\linewidth]{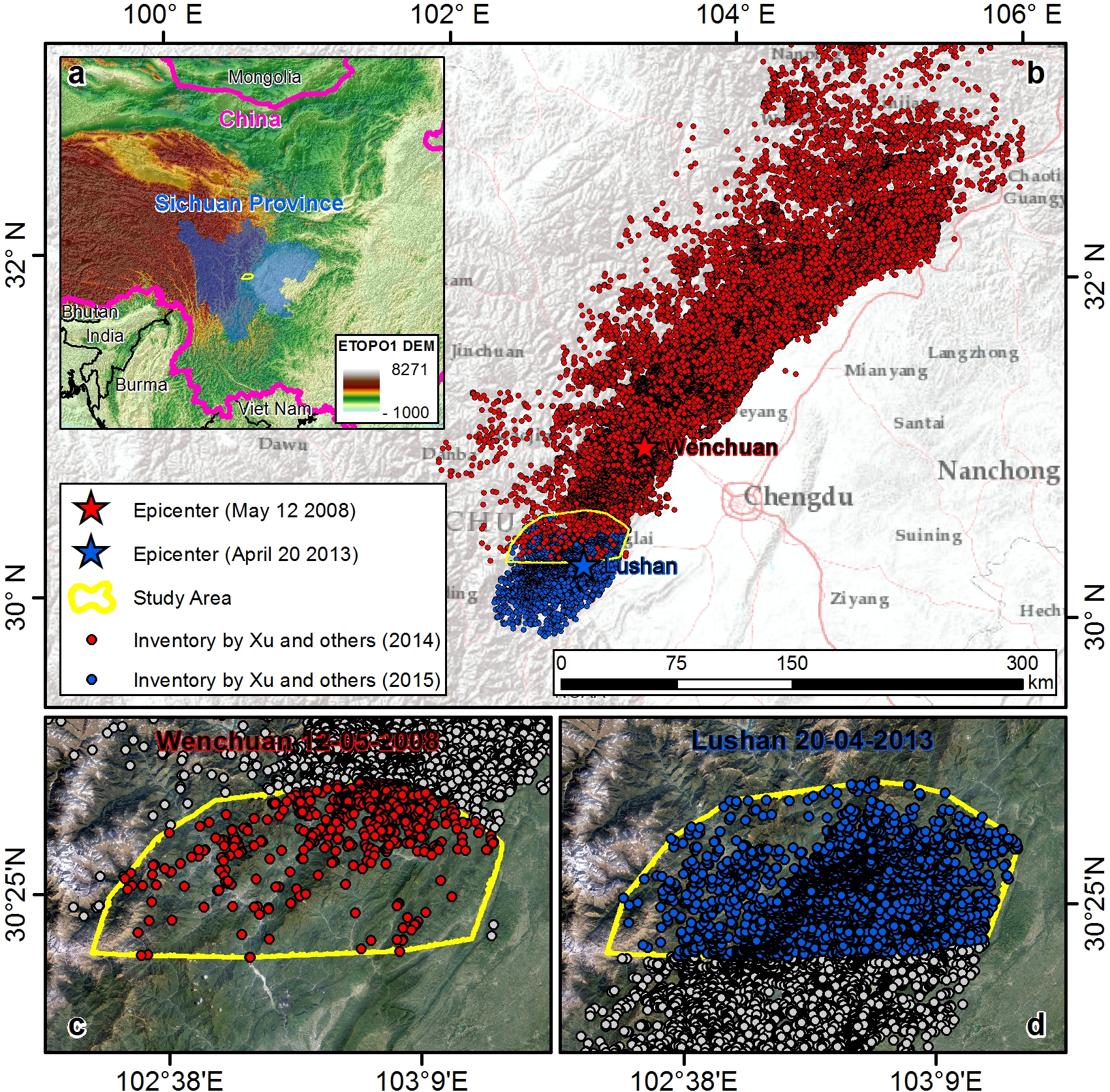}
	\caption{(a) Geographic context; (b) region hit by the two earthquakes; (c) and (d) landslide subsets based on the coinciding area.}
	\label{fig:map}
\end{figure}

We considered two earthquake-induced landslide scenarios in this work. In one case (Lushan), the inventory is close to the epicenter; in the other (Wenchuan), it is far. Thus, we tested our method on two distinct and extreme situations. The inventories were obtained from the first global earthquake-induced
landslide repository, which is developed and maintained by the U.S. Geological Survey and partners \citep{Schmitt2017,Tanyas2017}.

To build separate statistical models based on the Lushan \citep{Xu2015} and Wenchuan \citep{XU2014} inventories (see Figure \ref{fig:map}b), we initially digitized two polygons, each one encompassing all the landslides caused by each earthquake. Subsequently, we intersected the area between the two polygons (see Figure \ref{fig:map}c,d) and extracted all the landslides contained in the intersection. In the case of the Lushan earthquake, the inventory was provided as landslide centroids in vector format. For the Wenchuan earthquake, the inventory consisted of polygons encompassing the whole landslide scar, from the source to the deposition areas. Therefore, a polygon-to-point conversion was required for the Wenchuan inventory. In the literature, two conversion methods are available: \textit{i}) extracting the highest location along the perimeter of the landslide scar \citep[e.g.,][]{cama2015predicting,lombardo2016a}; or \textit{ii}) computing centroids for each polygon \citep[e.g.,][]{Hussin2016,Zerere2017}. Here, we selected the centroid option for consistency between the Lushan and Wenchuan datasets. 

The Lushan subset inventory contains $7868$ landslides, whereas the Wenchuan subset includes only $928$ landslides. We investigated whether any interaction existed between the hillslopes where slope failures occurred during both the earlier Wenchuan and the later Lushan earthquakes, which could indicate landslide reactivations. From this preliminary assessment, we found that less than $1\%$ of the landslides occupied the same slopes, suggesting that the signal in the Lushan data due to reactivation is negligible.

\subsection{Covariate Selection}
\label{sec:Covariates}

From a $90$~m digital elevation model (DEM) we derived the following morphometric covariates: \textit{i}) \emph{Elevation}; \textit{ii}) \emph{Slope} \citep{zevenbergen1987quantitative}; \textit{iii}) \emph{Eastness} and \emph{Northness} (i.e., the sine and cosine of the Aspect, respectively) \citep{Lombardo2018}; \textit{iv}) \emph{Planar} and \emph{Profile Curvatures} \citep{heerdegen1982quantifying}; \textit{v}) \emph{Relative Slope Position} \citep{bohner2006spatial}; \textit{vi}) \emph{Topographic Wetness Index} \citep{beven1979physically}; and \textit{vii}) \emph{Landform Classes} \citep{weiss2001topographic}. We also computed the Euclidean distances from each pixel to the nearest fault line, stream, and geological boundary to produce \emph{Distance to Faults}, \emph{Distance to Streams}, and \emph{Distance to Geoboundaries}. In addition to these geomorphic properties, we considered \emph{Outcropping Lithology} \citep{Ding2015}, \emph{Land Cover} \citep{sayre2014}, and \emph{Average Temperature Difference}. To calculate \emph{Average Temperature Difference}, we used Global Climate Data \citep{Fick2017} which report minimum and maximum temperatures for each month at every pixel. Then we computed the difference between the maximum and minimum temperatures for each month and their average per year. These covariates are shared between the two Wenchuan and Lushan intensity models, leaving specific differences only as a response to the shaking levels experienced across the landscape. We opted for such a large covariate set to account for most of the preconditioning factors known to contribute to slope instabilities. Therefore, the missing influence on the spatial patterns of landslide initiation should only be related to the actual shaking of the earthquake, which is represented by ground motion at each pixel.

We incorporated the spatial signals of the Wenchuan and Lushan earthquakes into the models by separately considering several ground-shaking parameters known for their impact in seismically-induced landslide hazard models: \textit{i}) \emph{Distance to Epicenter} (computed as the Euclidean distance from every pixel in the area to the epicenters of the two earthquakes); \textit{ii}) \emph{Macroseismic Intensity} \citep[MI,][]{kritikos2015regional}; \textit{iii}) \emph{Peak Ground Acceleration} \citep[PGA,][]{Nowicki2014}; and iv) \emph{Peak Ground Velocity} \citep[PGV,][]{Jessee2017}. However, these parameters do not directly express the frequency and duration of the ground motion \citep{Allstadt2018}, factors that control the landslide initiation process \citep[e.g.,][]{Jibson2004,Jibson2011}. Therefore, we also considered spectral acceleration maps via v) \emph{Peak Spectral Acceleration} at 0.3 s, 1.0 s, 3.0 s (PSA03, PSA1, and PSA3). Together these parameters reflect the peak response of a single degree of freedom oscillator; however, they also carry some information about the dominant frequency of an earthquake. All these shaking parameters, both for the Wenchuan\footnote{\url{https://earthquake.usgs.gov/earthquakes/eventpage/usp000g650\#shakemap}} and Lushan\footnote{\url{https://earthquake.usgs.gov/earthquakes/eventpage/usb000gcdd\#shakemap}} events, were collected from the U.S. Geological Survey (USGS) ShakeMap Atlas 2.0 \citep{garcia2012shakemap}.

\subsection{Mapping Units}
\label{sec:Units}

We considered four different mapping units: pixels, slope units, catchments, and administrative boundaries. The pixels correspond to a square lattice with $90$~m sides that coincides with the DEM and represent both the level of resolution at which the statistical model was built and the spatial partition over which we produced the reference landslide intensity map.

We computed the slope units via \textit{r.slopeunits} \citep{alvioli2016automatic} using the following parameterization: \textit{i}) initial flow~accumulation = $8\times 10^5~{\rm m}^2$; \textit{ii}) reduction factor = $2$; \textit{iii}) minimum slope unit area = $5\times 10^4~{\rm m}^2$; \textit{iv}) minimum circular variance = $0.35$; \textit{v}) clean area = $25\times 10^3~{\rm m}^2$; and \textit{vi}) number of iterations = $10$. Due to the large size of the study area, we opted for a large flow accumulation to limit the overall number of slope units. In our statistical models, we defined the latent spatial effect over the slope units, and we aggregated the landslide pixel intensity over these spatial units. The role of the pixels and slope units in our landslide intensity model is described in \S\ref{sec:strategy}.

Ultimately, we used the catchments and administrative boundaries only as spatial partitions to project the computed intensities. We demonstrate this in the next section, \S\ref{sec:Poisson}.

\section{Point Processes for Landslide Intensity Modeling}
\label{sec:popland}

\subsection{Using a Point Process Instead of the Bernoulli Distribution}
\label{sec:Poisson}
The presence and absence of events (such as landslide occurrences) are often modeled using a Bernoulli distribution. In spatial modeling, however, it is not trivial to make the Bernoulli distribution consistent across spatial resolutions, i.e., different Bernoulli models would have to be fitted for different pixel sizes or mapping units \citep{cama2016exploring}. 
For example, when we model landslides at two resolutions, a $30$~m~$\times$~$30$~m and a $60$~m~$\times$~$60$~m, the two resulting Bernoulli logit models behave in fundamentally different ways.
The first logit model produces four occurrence probabilities for each probability produced by the second model; summing these four probabilities (or computing the probability that at least one of the four has a positive occurrence) does not yield the same result as the second model. 

In contrast, the Poisson distribution is consistent across all spatial resolutions.
This is due to the property of Poisson additivity, i.e., that the sum of $N$ independent Poisson variables with mean $\lambda_i$, $i=1,\ldots, N$ is again a Poisson variable with mean $\sum_{i=1}^N \lambda_i$. Using this Poisson additivity, we can define a spatial Poisson process over a continuous space such that each region $A$ contains a random number of events (e.g., landslides) that follow the Poisson distribution with mean $\lambda(A)=\int_A\lambda(s){\rm d}s$, where $\lambda(s)\geq0$ denotes the intensity at location $s$. Essentially, the intensity $\lambda(s)$ represents the ``density'' of events around location $s$, while the integrated intensity $\lambda(A)$ is the expected number of events occurring in the set $A$. For larger sets $B\supseteq A$, $\lambda(B)\geq\lambda(A)$ (in other words, the expected number of events in $B$ is larger than that in $A$). A spatial Poisson process with a spatially varying random intensity $\lambda(s)$ is called a Cox process, and if $\lambda(s)$ is modeled as a Gaussian process on the log scale, then it is known as a log-Gaussian Cox process (LGCP). Unlike Poisson processes, LGCPs allow for the incorporation of all sorts of fixed effects (i.e., continuous covariates) and categorical, spatial or temporal random effects (e.g., the LSE) into the intensity function $\lambda(s)$. They are widely used in geostatistics to model point patterns that exhibit spatial clustering characteristics.  
If the intensity is constant over an area, then the expected number of landslides in a $60$~m~$\times$~$60$~m grid cell is four times the expected number in a $60$~m~$\times$~$60$~m grid cell, because the area is four times larger. This also applies to very low intensities in a small area, where the probability of occurrence is still approximately four times higher in the larger grid cell than the smaller.
Therefore, the models with different resolutions are directly comparable, and we can sum the intensity over any number of grid cells to calculate the intensity in a given mapping unit.

We considered an LGCP \citep{Simpson.al.2016} to capture the landslide patterns caused by seismic shaking, where the intensity is defined as $\lambda(s) = e^{\eta(s)}$ and $\eta(s)$ is modeled as a Gaussian process characterized by fixed covariate effects and random effects, as described below in \S\ref{sec:strategy}. We defined this LGCP on a gridded space, assuming that the intensity function $\lambda(s)$ was constant within each pixel. This dramatically simplified the integrals to compute, but was still accurate for small pixel sizes.

The implied Bernoulli probability of presence, i.e., the susceptibility, in an area is the probability that the Poisson probability is greater or equal to 1; in other words,
\begin{align}
\text{Susceptibility} = 1-e^{-\lambda_A}, 
\label{eq:conversion}
\end{align}
where $\lambda_A=\sum_{i\in A}\lambda_i$ is the summed intensity of pixels in area $A$.
This formula, which we refer to as the intensity-to-susceptibility conversion, allows one to easily and quickly compute the susceptibility for any mapping unit $A$ of interest from the corresponding pixel intensities.

\subsection{Model Building Strategy}
\label{sec:strategy}
We numbered the grid cells (pixels) as $s_i$, $i=1,\ldots,N$, writing the $i$th pixel intensity as $\lambda_i=\exp(\eta_i)=\exp\{\eta(s_i)\}$. We modeled $\eta(s)$ as a Gaussian process in terms of the sum of fixed and random effects, which for the $i$th pixel has the form
\begin{align}\label{eq:linearpredictor}
\eta_i = \beta_0 + X_i \beta + u_i + v_{1, i}+ v_{2, i} + \cdots ~.
\end{align}
Here, $X_i$ denotes the vector of linear covariates for the $i$th pixel, $\beta$ is the corresponding vector of fixed effects, $u_i$ is a spatial random effect, and $v_i$s are random effects for the categorical covariates. All these variables have Gaussian priors, conditionally on a small set of hyperparameters \citep{Rue2009,Rue.al.2017}. Each additive term in \eqref{eq:linearpredictor} is a model component, capturing the effects that covariates have on the log-intensity.

There are many models that can be considered as alternatives to the LSE, which is described by the vector $\boldsymbol{u}=(u_1, \ldots,u_N)^T$. We followed \citet{Lombardo.etal:2018} and assumed pixels within the same slope unit to have a much stronger spatial dependency than pixels from different slope units (even when they are closer to each other in Euclidean distance). We numbered the slope units as $j=1,\ldots,M$, and identified the set of pixels within the $j$th slope unit as $I_j\subset\{1,\ldots,N\}$. To define the LSE on the slope units themselves, $\boldsymbol {U}=(U_1,\ldots,U_M)^T$ denotes the vector of spatial effects for the slope units, and the correspondence with pixels is defined as $u_i=U_j$ for all $i\in I_j$. We used an LSE based on a Besag model (also known as an intrinsic conditional autoregressive model, or iCAR model, see \citet{besag1991bayesian}), which may be expressed in terms of conditional distributions as follows:
\begin{align}\label{eq:iCAR}
  U_j \mid \boldsymbol{U}_{-j}, \tau_0 \sim \mathcal N \left( \frac{1}{d_j} \sum_{k \sim  j} U_k,
  \frac{1}{d_j} \frac{1}{\tau_0} \right),
\end{align}
$\boldsymbol{U}_{-j}$ denotes the vector $\boldsymbol{U}$ without the $j$th element, $k \sim j$ signifies that the $k$th and $j$th slopes units are neighbors, and $d_j$ is the number of neighbors of the $j$th slope unit.  The precision parameter, $\tau_0>0$, controls the overall importance (i.e., the ``size'') of the LSE in \eqref{eq:linearpredictor}. Smaller values of $\tau_0$ imply that the LSE is more important. For identifiability reasons, we added a sum-to-zero constraint on $\boldsymbol {U}$ and rescaled the model as detailed by \citet{sorbye2014scaling}. Essentially, the Besag model \eqref{eq:iCAR} assumes that the spatial effect for a specific slope unit is only related to the slope unit through its direct neighbors, inducing spatial correlation.

We split the \textit{Macroseismic Intensity} (MI) covariate into $20$ equidistant classes, and assumed that a first-order random walk drives the dependence structure among the corresponding class effects $v_{1,1},\ldots,v_{1,20}$. That is, we assumed  
\begin{align}
v_{1, i} &= v_{1, i-1} + e_i,\qquad e_i\sim \mathcal N(0, \tau_1^{-1}),
\end{align}
where $\tau_1>0$ is the corresponding precision parameter. This random walk model is useful for capturing potential non-linear effects of the MI covariate. As with the Besag model, we added a sum-to zero constraint and rescaled the model.

The covariates \emph{Land Cover}, \emph{Landforms}, and \emph{Lithology} are categorical in nature, with no particular structure among classes, so we modeled them using independent random effects. That is, for each categorical covariate $k\geq 2$,
\begin{align}
v_{k, i} \iid \mathcal N(0, \tau_k^{-1}),
\end{align}
and $\tau_k>0$ is the corresponding precision parameter. 

For each of the hyperparameters $\tau_0$, $\tau_1$, and $\tau_k$, $k\geq 2$, we assumed a penalized complexity prior \citep{simpson2017penalising} on $\sigma_k = \tau_k^{-1/2}$, which is an exponential prior with median at $\sigma_k =0.1$. This prior allowed us to maintain high flexibility and to estimate the hyperparameters, and prevent overfitting by shrinking this complex model towards a simpler one.

\subsection{Bayesian Inference Using INLA}
In the Bayesian methodology, after a prior model (which can be sampled from) is defined and a dataset $\vec y$ obtained, the posterior distribution can be computed according to the basic laws of probability.
However, performing this computation numerically can be quite challenging.
We exploited the integrated nested Laplace approximations (INLA) package in R to make inference and efficiently compute the posterior distributions.
A general review of INLA can be found in \citet{Rue.al.2017}, while a more detailed review of spatial modeling can be found in \citet{Bakka.etal:2018}.

INLA divides the model into three stages: an observation likelihood $\pi(\vec y \mid \vec \eta)$ (Poisson distribution), a linear predictor $\vec \eta=(\eta_1,\ldots,\eta_N)^T$ at the pixel level, and a set of hyperparameters $\vec \theta$.
Following the notation from \cite{Bakka.etal:2018}, the posterior distribution for hyperparameters may be expressed as
\begin{align}
\pi(\vec \theta \mid \vec y) \propto \frac{\pi(\vec y \mid \vec \eta= \vec a, \vec \theta) \pi(\vec \eta= \vec a \mid \vec \theta)}{\pi(\vec \eta = \vec a \mid \vec y, \vec \theta)} \pi(\vec \theta), \label{eq-laplace2}
\end{align}
where $\pi(\vec \eta = \vec a \mid \vec y, \vec \theta)$ is approximated by a Gaussian distribution, and
the mode $\vec a$ is found for each value of $\vec \theta$ by optimizing 
$\pi(\vec \eta = \vec a \mid \vec y, \vec \theta)$ iteratively.
INLA applies a version of the gradient descent to find the maximum posterior, 
\begin{align}
\vec {\hat {\theta}} = \text{argmax}_{\vec \theta} \  \pi(\vec \theta \mid y),
\end{align}
which is then used in the computation of $\pi(\vec \eta\mid \vec y, \vec\theta)$.

\subsection{Performance Metrics for Dichotomous and Count Data}
\label{sec:Metrics}

In this paper, we make both within-sample and out-of-sample model comparisons.
The within-sample comparisons (measures of fit) aim to show how closely the models fit to the data. 
However, we can obtain arbitrarily good fits (overfitting) with complex models, rendering these measures unsuitable for comparing models with any interesting level of complexity.
To alleviate this, we performed ten-fold cross-validation (CV), 
where we uniformly divided the dataset into ten subsets at random, 
estimate the model from nine of the subsets, and predict on the last subset. These ten subsets are constrained to be complementary; in other words, their union returns the original dataset.

For the binary interpretation, we compared point predictions of presence probability (using the estimated pixel intensities as in \eqref{eq:conversion}) to the true presence-absence via the receiver operating characteristic (ROC) curve \citep{hosmer2000}. We chose the ROC curve because it has a long history in landslide susceptibility modeling \citep[e.g.,][]{Frattini2010,Pourghasemi.etal:2013}, and because it is among the most common metrics for binary data \citep[e.g.,][]{Hong2018,Tziritis2017}.
The strength of this metric is that it represents the modeling goal very well at the pixel level (most counts are 0 and 1).
Its weakness is that it does not represent the data well at aggregated levels, where counts may vary significantly.
To summarize the ROC curve, we used the area under the curve (AUC) and followed the interpretation proposed by \citet{hosmer2000}, where the model performance is classified as follows: \textit{i}) $0.7 < \mbox{AUC} < 0.8$: acceptable; \textit{ii}) $0.8 < \mbox{AUC} < 0.9$: excellent; and \textit{iii}) $0.9 < \mbox{AUC} < 1.0$: outstanding results.

Next, we interpreted the count data, which represents the number of landslides in certain mapping units. We compared predictions of the expected number of counts over the mapping unit $A_k$ (i.e., the intensity $\lambda(A_k)$), which are estimated from the model, to the true number of landslides observed in that area, denoted by $Y(A_k)$.
To summarize the agreement between $\lambda(A_k)$ and $Y(A_k)$, we computed two additional metrics.
For the first metric, similar to the common ``R-squared'' metric, we defined R$2$ as
\begin{align}
\mbox{R2} = 1-\frac{\text{Var}_k \{Y(A_k)-\lambda(A_k)\}}{\text{Var}_k \{Y(A_k)\}},
\end{align}
representing the percentage of variability in the counts $\{Y(A_k)\}$ explained by their model-based counterparts $\{\lambda(A_k)\}$. For the second metric, we introduced the ratio of explained counts (RCE), 
\begin{align}
\mbox{RCE} = 1- \frac{\sum_k |Y(A_k)-\lambda(A_k)|}{\sum_k Y(A_k)},
\end{align}
where the denominator always equals the total number of observed counts in the dataset.
The RCE can be interpreted as the percentage of landslides that are identified or predicted in a specific mapping unit compared to the original count data. Note that an in-sample RCE of near 1 indicates a near perfect fit (overfitting).

\subsection{Seismic variable selection}
\label{sec:Varselection}

Statistical models should avoid multicollinearity in the data. Because all the shaking parameters for both the Lushan and Wenchuan earthquakes came from the same source, we investigated potential linear correlations among the covariates. Supplementary Figures SM1 and SM2 report the Pearson correlation coefficients of all the continuous covariates, and a strong dependence can be seen among \emph{Distance to Epicenter}, PGA, PGV, MI, PSA03, PSA10, and PSA30 for both earthquakes. Therefore, only one of these seven covariates should be included in the model. We used the intensity-to-susceptibility conversion formula in \eqref{eq:conversion} and converted the counts to binary presence-absence data in order to calculate AUC values (Table \ref{table1}) from point process models based on single explanatory variables. Macroseismic Intensity (MI) appears to be the best covariate overall to carry the  earthquake signal into our subsequent models. MI may perform best as an earthquake-related parameter because it combines both the amplitude of shaking and the shaking duration, which both depend on distance, into an aggregate measure of the earthquake effect.

\begin{table}[t!]
\vspace{5pt}
\centering
\begin{tabular}{c|c|c|c|c|c|c|c} 
& \textbf{Distance to Epicenter} & \textbf{MI} & \textbf{PGA} & \textbf{PGV} & \textbf{PSA03} & \textbf{PSA1} & \textbf{PSA3} \\ 
\hline
Lushan & 0.763 & 0.783 & 0.785 & 0.782 & 0.785 & 0.763 & 0.770 \\
Wenchuan & 0.821 & 0.823 & 0.804 & 0.790 & 0.713 & 0.750 & 0.745 
\end{tabular}
\caption{Model predictive performance in terms of AUC by using single seismic variable point process models.\label{table1}}
\end{table}

\section{Results}
\label{sec:results}

\subsection{Capturing Shaking Levels via the Latent Spatial Effect}
\label{sec:LSE}

\begin{figure}[t!]
\centering
\includegraphics[width=0.9\linewidth]{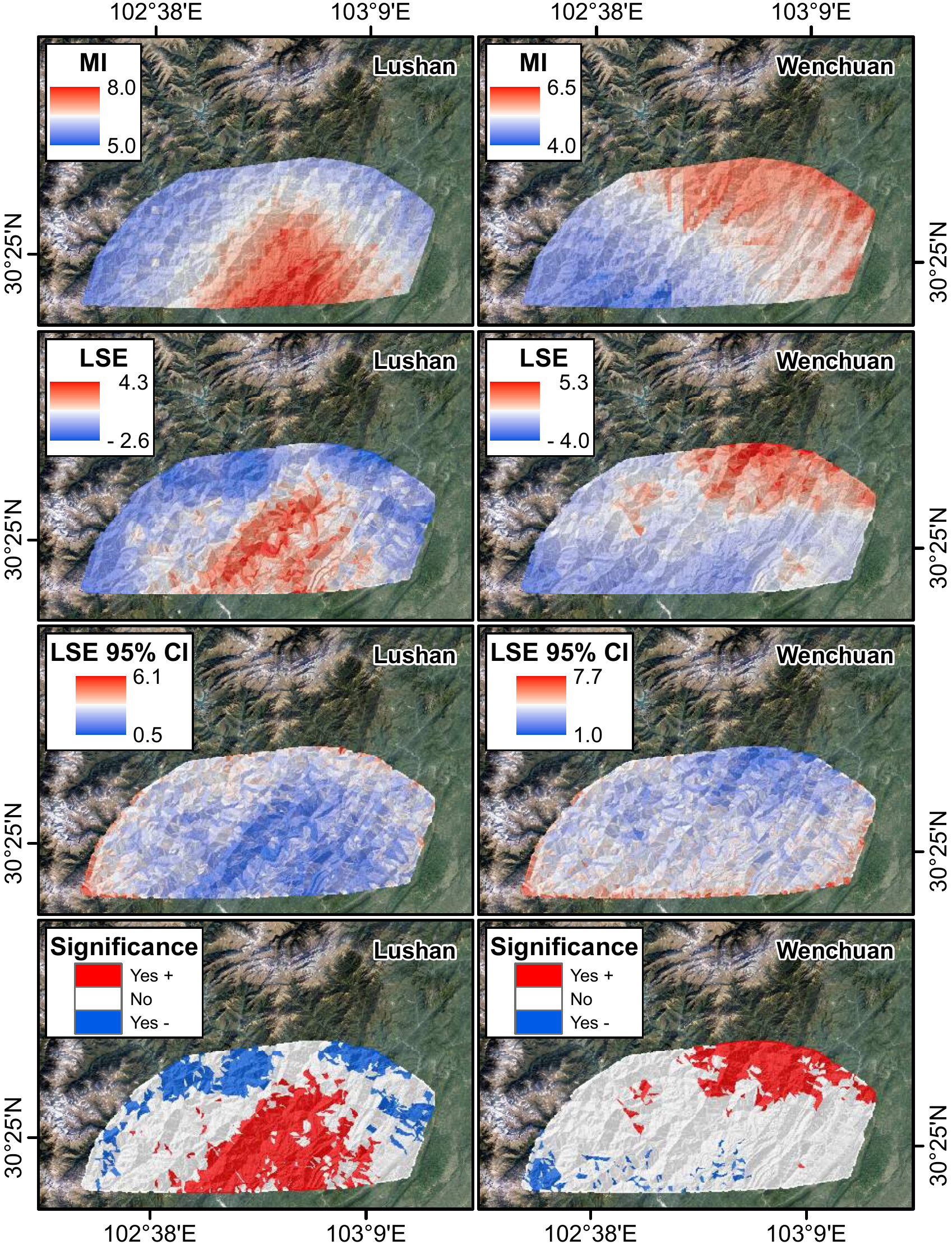}
\caption{First and second rows: Comparison between the spatial patterns of the MI covariate and the posterior mean of the LSE. Third and fourth rows: Uncertainty and significance of LSE. Columns: Lushan (left) and Wenchuan (right) earthquakes.}
\label{fig:LSE_VS_MI}
\end{figure}

We fitted two models with the LSE (without MI) to the two datasets and obtained one LSE for each of the two earthquakes. Figure \ref{fig:LSE_VS_MI} shows the two MI covariates compared to the posterior mean of the LSEs together with their $95\%$ credible intervals and significance. The credible interval is defined here as the difference between the 0.975 and 0.025 quantiles of the LSEs, and the significance is considered as the slope units where these quantiles have the same sign. 

\begin{figure}[t!]
\centering
\includegraphics[width=0.6\linewidth]{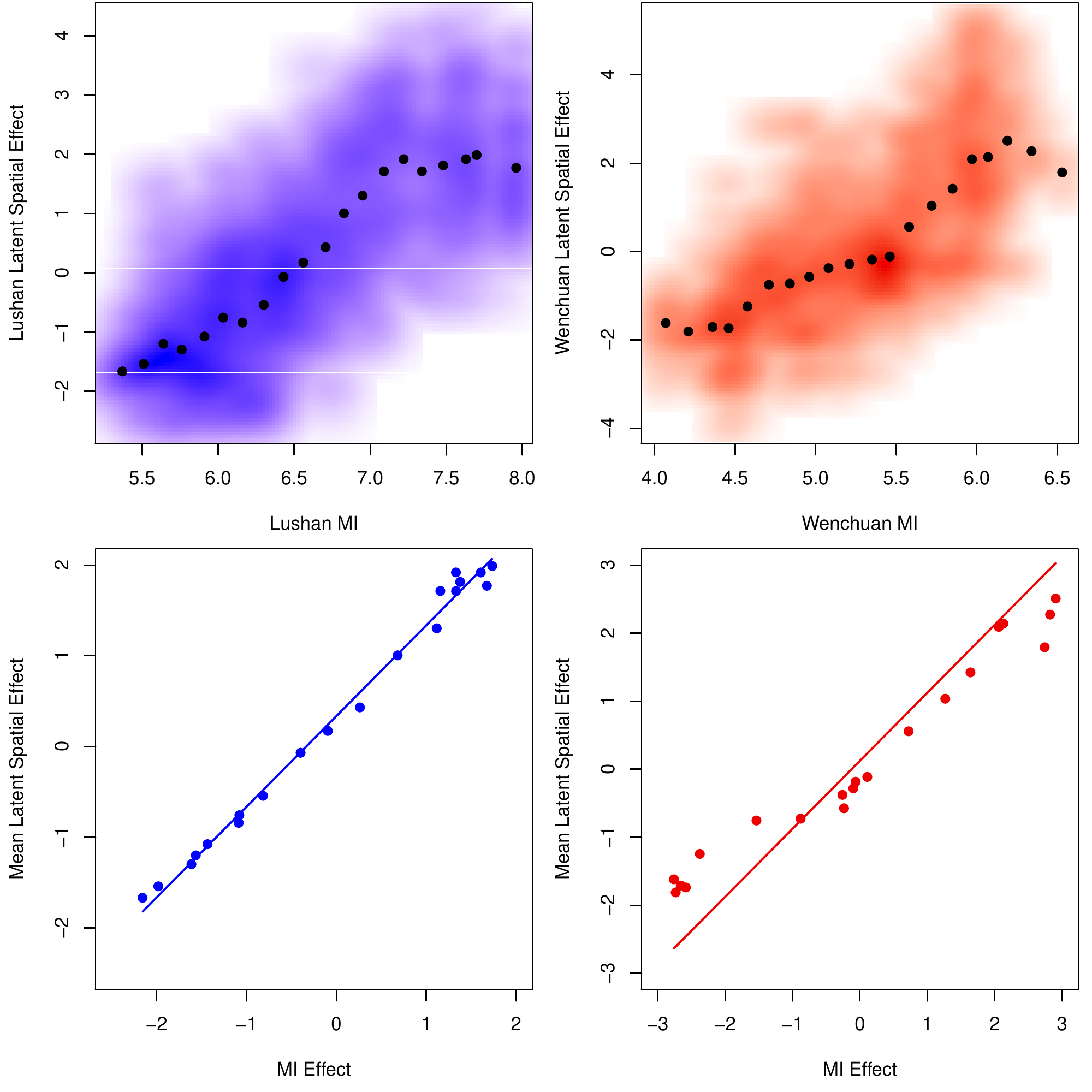}
\caption{Quantitative comparison between LSE and MI for Lushan and Wenchuan earthquakes. The first row shows a 2D kernel density scatterplot of the original MI map against the estimated LSE map. Black dots are the average LSE calculated for each of the 20 nonlinear classes of the MI used in the MI-only model. The second row shows the average LSE for each MI class plotted against the actual coefficients obtained for each of the 20 classes in the MI-only model.}
\label{fig:ComparisonLSEandMI}
\end{figure}

The patterns shown in Figure \ref{fig:LSE_VS_MI} provide a visual comparison between the shaking parameters and the LSEs of the two earthquakes; Figure \ref{fig:ComparisonLSEandMI} presents a more quantitative comparison between the MI and the estimated LSE. This is evaluated both between the LSE and the original form of the MI as a covariate. We also investigated the relation between the LSE and the effects that MI, i.e., its coefficients, have within the MI-only model for both earthquakes.

\begin{figure}[t!]
\centering
\includegraphics[width=0.8\linewidth]{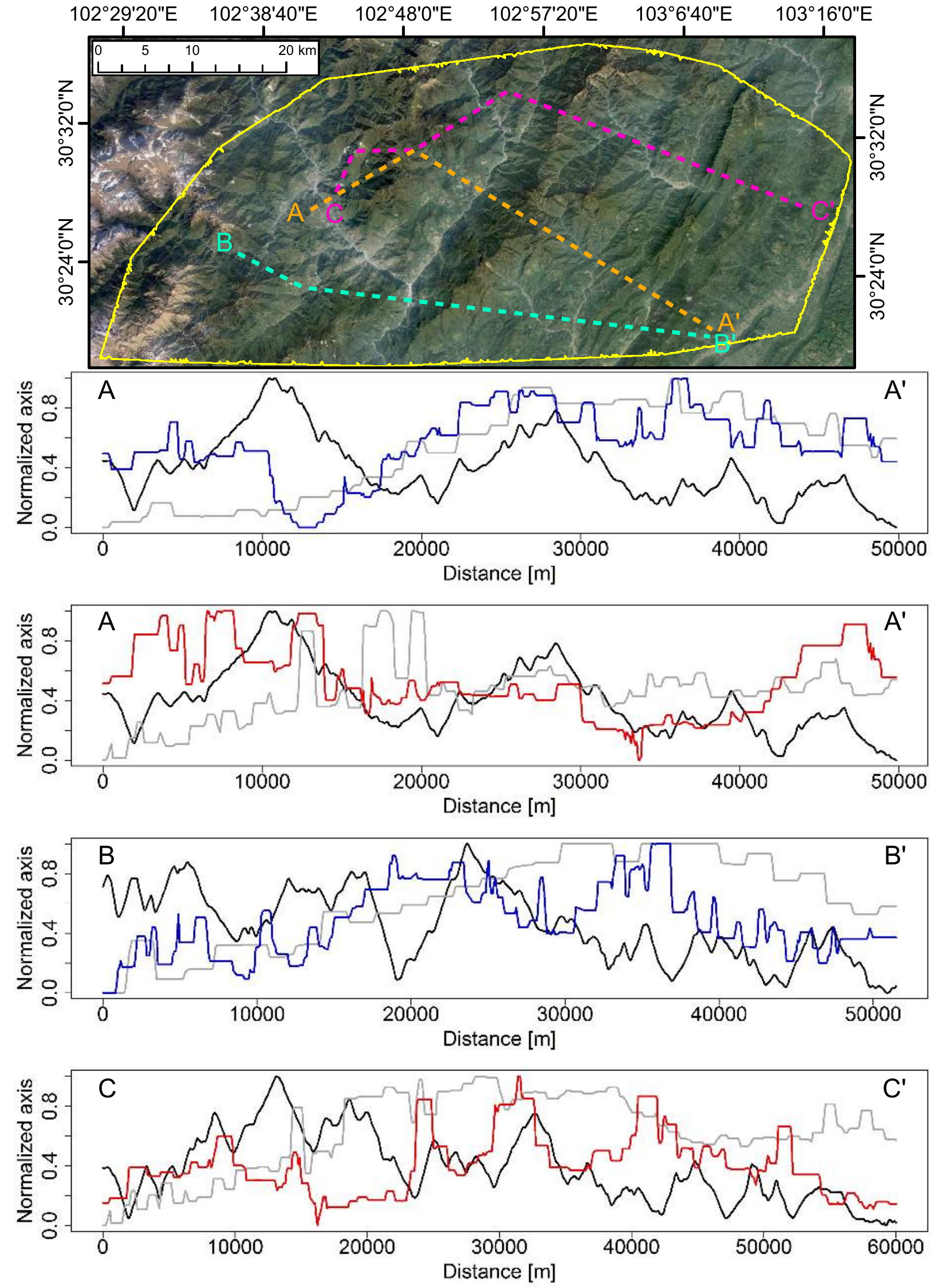}
\caption{Profiles of elevation (black), MI (grey), and estimated LSE (blue for Lushan and red for Wenchuan models) along different cross-sections A-A', B-B' and C-C' illustrated on the map.}
\label{fig:CrossSections}
\end{figure}

To complete the comparison between the shaking parameters and the LSE, we generated three cross sections (see Figure \ref{fig:CrossSections}): \textit{i}) Section A-A', which passes between the areas where Lushan and Wenchuan produced the highest shaking and is used for both earthquakes; \textit{ii}) Section B-B', which cuts through the landscape in the south where the Lushan earthquake occurred; and \textit{iii}) Section C-C', which captures the Wenchuan signal in the north. This representation aims at uncovering  potential topographic amplifications. The MI does not appear to follow any narrow peak or summit whereas the LSE, although not in every case, varies much more frequently with spikes that coincide with certain ridges.  

\subsection{Model Selection via Within-Sample Performance}
\label{sec:Assessing Performance}

We initially assessed the performance of our fitted models in terms of binary presence-absence responses. The estimated model-based intensities were transformed into classic susceptibility values (see \S\ref{sec:popland}, Equation \eqref{eq:conversion}), and the observed counts were converted into their binary counterparts. Thus, we computed both the ROC curves and their AUC values for consistency with traditional landslide susceptibility studies.

Table \ref{table2} reports the AUC values for each fitted model, for both the Lushan and Wenchuan earthquakes. Hence, we can be compare the performances of the MI-only model, LSE-only model, and a model with both MI and LSE. 
A clear pattern arises where the MI-only model is the weakest, the LSE-only model is the best, and the MI-and-LSE model performs no better than the simpler LSE-only model within the first three decimal places \citep[according to ][]{hosmer2000}. 
Therefore, for the remainder of this paper, we consider the LSE-only model for both the Lushan and Wenchuan earthquakes.

\begin{table}[t!]
\vspace{5pt}
\centering
\begin{tabular}{c|c|c|c} 
& \textbf{MI} & \textbf{LSE} & \textbf{MI + LSE} \\ 
\hline
\textbf{Lushan} & 0.846 & 0.889 & 0.889 \\
\textbf{Wenchuan} & 0.871 & 0.943 & 0.943
\end{tabular}
\caption{AUC values for three model configurations at the pixel level to assess fitting performance. \label{table2}}
\end{table}

\subsection{Mapping Landslide Intensities}
\label{sec:Mapping}

\begin{figure}[t!]
\centering
\includegraphics[width=0.9\linewidth]{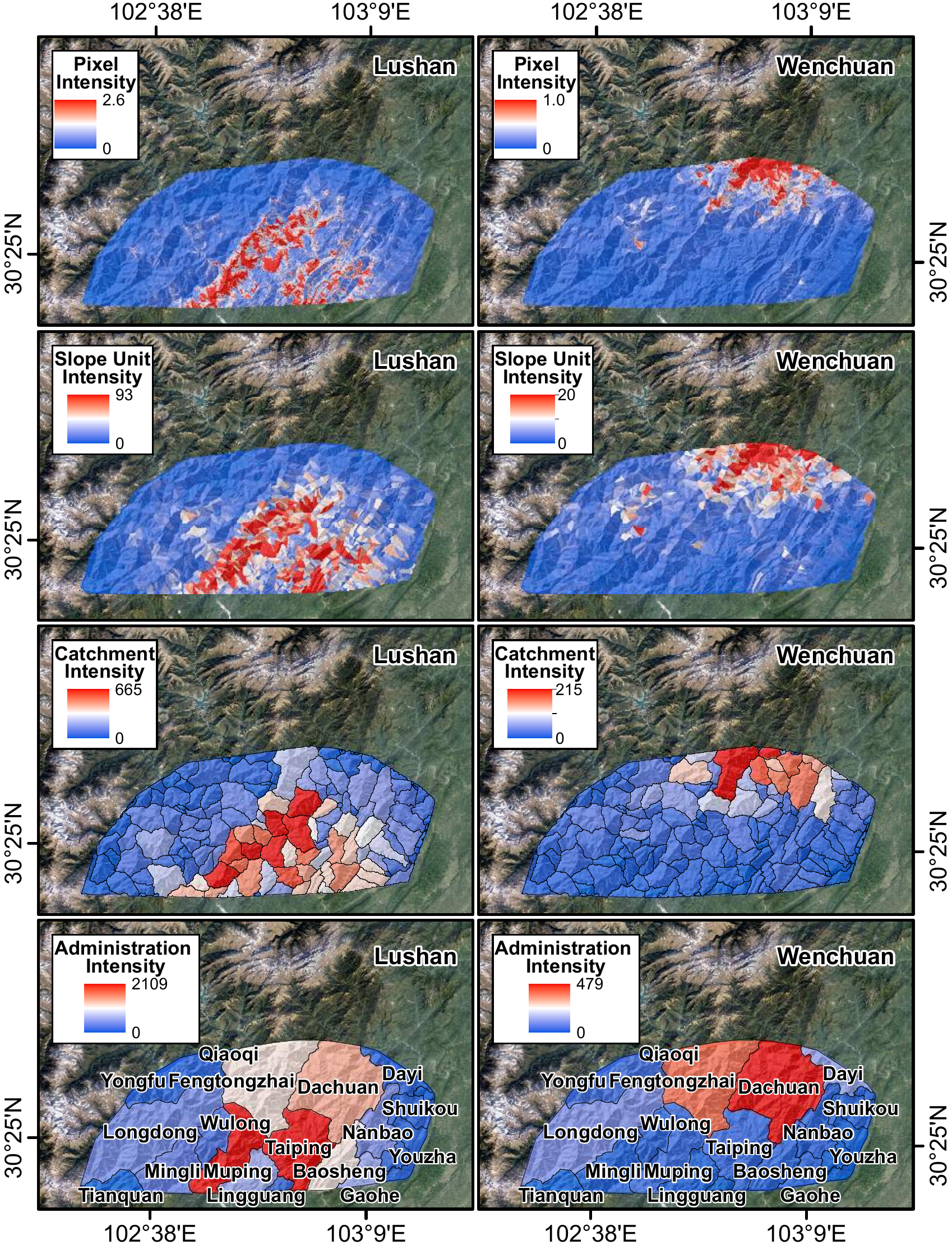}
\caption{Landslide intensity maps at pixel, slope unit, catchment, and administrative unit scales for both earthquakes.}
\label{fig:Intensity maps}
\end{figure}

The advantages of modeling intensities instead of susceptibilities using a suitable log Gaussian Cox process are that we can jointly predict \emph{how many} landslides and \emph{where} they will potentially occur. The aggregative property of intensity also allows us to generate predictive maps for any spatial unit using just one model (see \S\ref{sec:popland}). In this work, we considered four mapping units to illustrate the results: pixels, slope units, catchments, and administrative (sub-counties) units. We obtained the corresponding intensities from the initial pixel model and summed them over each polygonal object. Figure \ref{fig:Intensity maps} shows the landslide intensity maps generated for both earthquakes and all four mapping units.   

\subsection{Assessing Cross-Validation Performance}
\label{sec:Assessing Performance}

\begin{figure}[t!]
\centering
\includegraphics[width=0.4\linewidth]{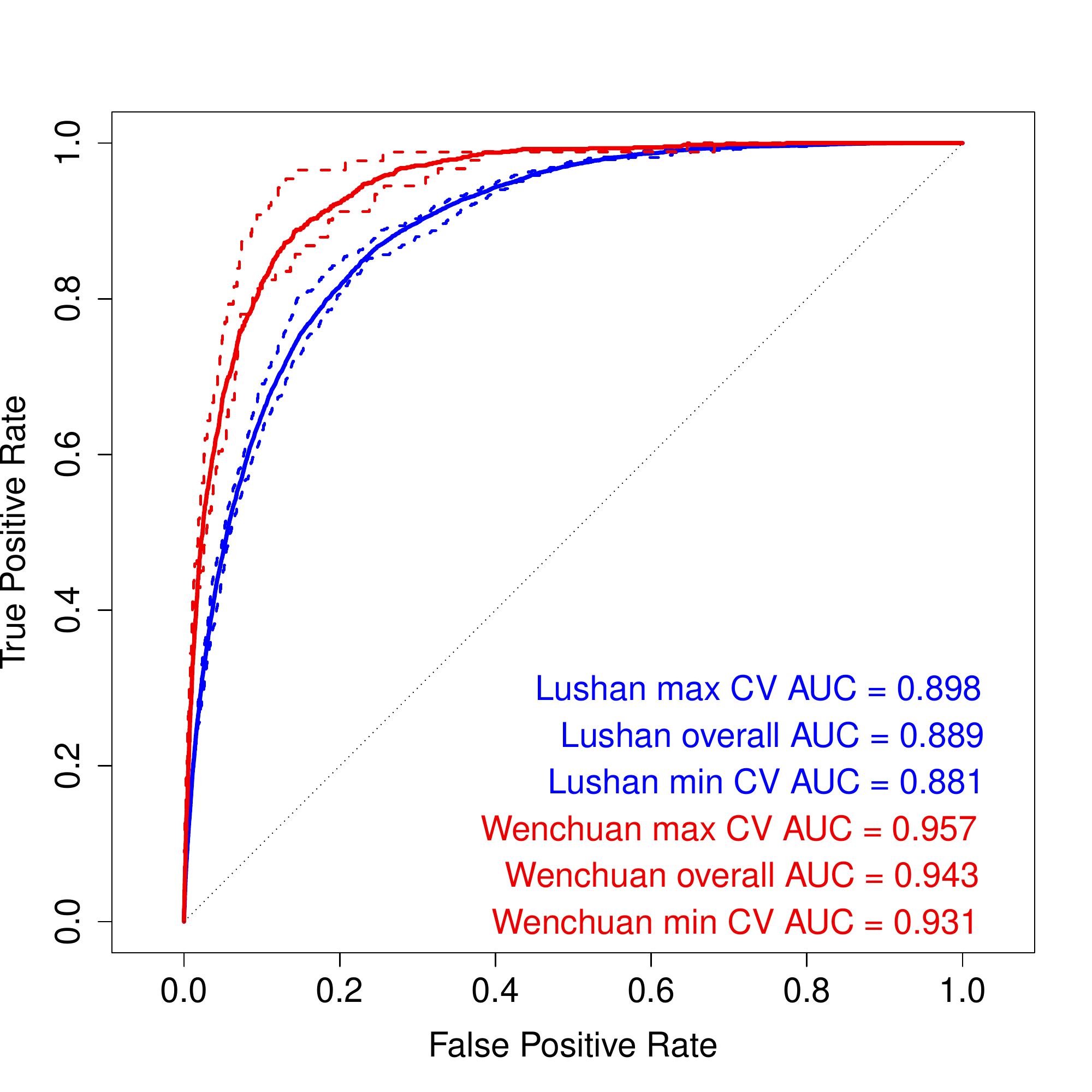}
\caption{ROC curves computed for the Lushan and Wenchuan earthquakes. Solid lines represent the overall ROC values using the entire dataset, whereas dashed lines show the maximum and minimum ROC curves obtained via the 10-fold CV.}
\label{fig:10foldCV}
\end{figure}

We computed several out-of-sample metrics, as described in Section \ref{sec:Metrics}, to assess the performance of the LSE-only models.
First, the ROC curves (Figure \ref{fig:10foldCV}) compare the predicted susceptibility to the observed presence-absence of landslides at the pixel level.
The solid line represents the overall out-of-sample ROC curve and the  dashed lines correspond to the minimum and maximum ROC curves (ranked by AUC) obtained from each of the ten CV folds.
We note that the overall ROC curves are contained within the most extreme out-of-sample ROC curves.

\begin{figure}[t!]
\centering
\includegraphics[width=\linewidth]{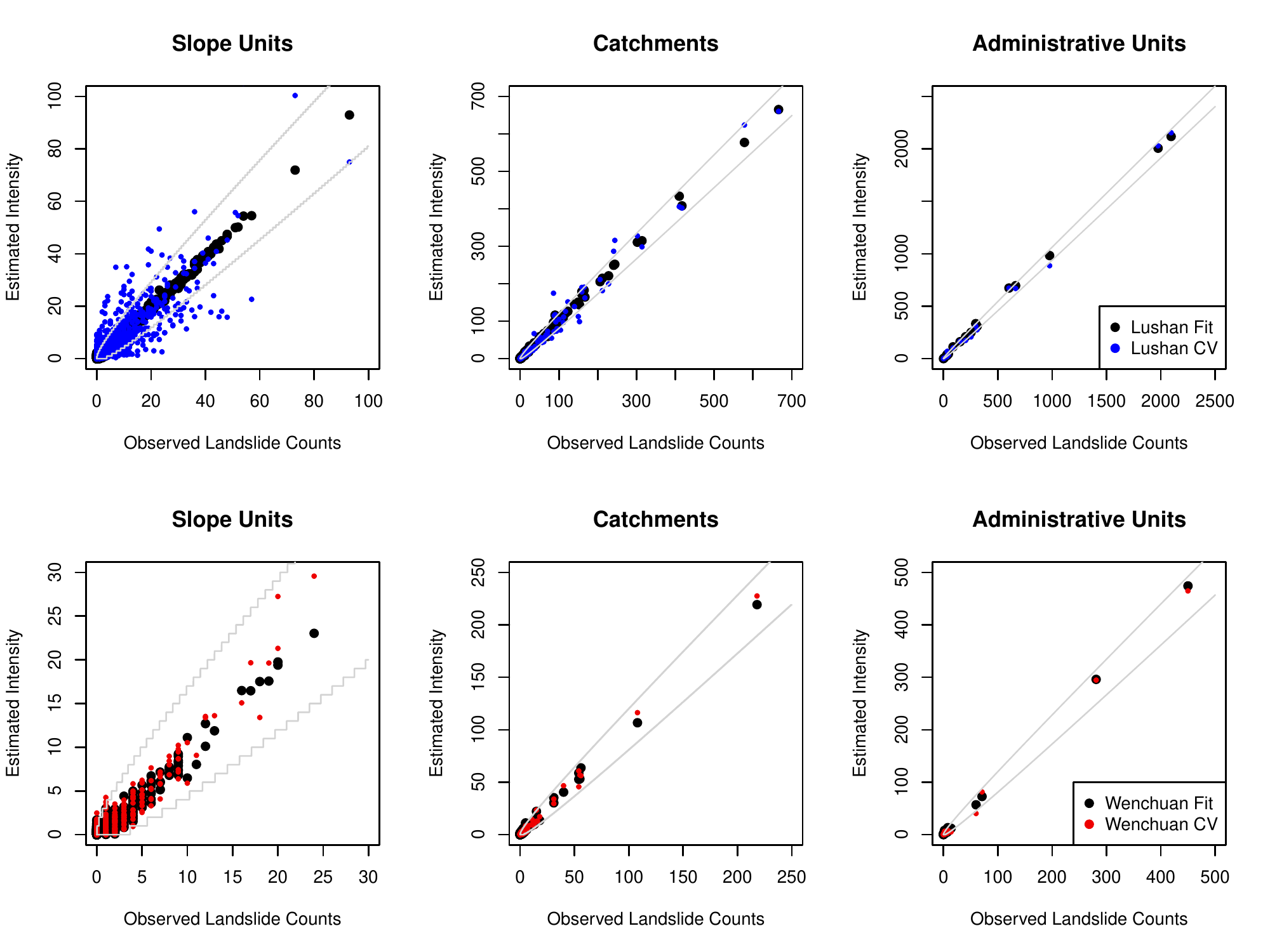}
\caption{Estimated landslide intensities against observed landslide counts for different mapping units. Black dots show the fitted values while colored dots correspond to CV results. Grey lines represent the theoretical 95\% credible intervals that would arise from a Poisson distribution with mean equal to the counts shown in the abscissas and ordinates.}
\label{fig:Counts_VS_Intensities}
\end{figure}

The landslide count data at the slope unit, catchment, and administrative unit scales are presented in Figure \ref{fig:Counts_VS_Intensities}.
Here, the observed landslide counts are compared to the estimated intensity (both within-sample and out-of-sample).
The fitted LSE-only models match the data extremely well, especially considering that the models are constructed at the pixel level, and the out-of-sample predicted counts have a limited spread around the observed ones.

Finally, the fitted models and CV results are summarized both in terms of landslide susceptibility and intensity over the four considered mapping units in Table \ref{table3}. The AUC values summarize the ROC curves, and the R2 and RCE summarize the performance of the models in the count domain (see \S\ref{sec:Metrics}). The three metrics hierarchically show an increasing performance. Although this is expected for susceptibility results, it highlights the strength of the intensity framework.    

\begin{table}[t!]
  \centering
  \renewcommand{\arraystretch}{1.2}
  \begin{tabular}{|l|c|c|c|c|c|c|c|c|c|c|}
    \hline
    \multirow{2}{0.2cm}{} & \multicolumn{1}{c|}{\textbf{Pixels}} & \multicolumn{3}{c|}{\textbf{Slope units}} & \multicolumn{3}{c|}{\textbf{Catchments}} & \multicolumn{3}{c|}{\textbf{Administrations}} \\
    \cline{2-11}
    & \textbf{AUC}  & \textbf{AUC} & \textbf{R2} & \textbf{RCE} & \textbf{AUC} & \textbf{R2} & \textbf{RCE} & \textbf{AUC} & \textbf{R2} & \textbf{RCE} \\
    \hline
    FIT1 & 0.889 & 0.951 & 0.987 & 0.808 & 0.985 & 0.998 & 0.944 & 1.000 & 0.999 & 0.996 \\ \hline
    CV1 & 0.850 & 0.916 & 0.727 & 0.484  & 0.981 & 0.977 & 0.860 & 1.000 & 0.966 & 0.934 \\ \hline
    FIT2 & 0.943 & 0.959 & 0.987 & 0.477  & 0.949 & 0.988 & 0.840 & 0.989 & 0.999 & 0.918  \\ \hline
    CV2 & 0.928 & 0.940 & 0.902 & 0.472 & 0.949 & 0.992 & 0.841 & 0.967 & 0.996 & 0.907  \\ \hline
  \end{tabular}
  \caption{Performance of the LSE-only model for all considered mapping units. Traditional performance metrics (AUC) for dichotomous data are reported together with our two proposed count performance metrics (R2 and RCE). FIT1 and CV1 refer to the within-sample and out-of-sample metrics for the Lushan earthquake, respectively; FIT2 and CV2 are their counterparts for the Wenchuan earthquake. \label{table3}}
\end{table}

\subsection{Covariate Effects}
\label{sec:Covariate Effects}

\subsubsection{General Overview}
\label{sec:general}

Our LSE-only model includes both \emph{fixed} (linear) and \emph{random} (nonlinear) effects. The following subsections will present both cases, including a novel representation of the contribution of Eastness and Northness to the model. 

Each plot shown in the following subsections reports the regression coefficients for each of the covariates in the model. The covariates were rescaled by subtracting the mean and dividing by the standard deviation, thus making each regression coefficient comparable to the others.

The geomorphological interpretation of these effects is provided in the Supplementary Material. 

\subsubsection{Fixed Effects}
\label{sec:Fixed Effects}

\begin{figure}[t!]
\centering
\includegraphics[width=\linewidth]{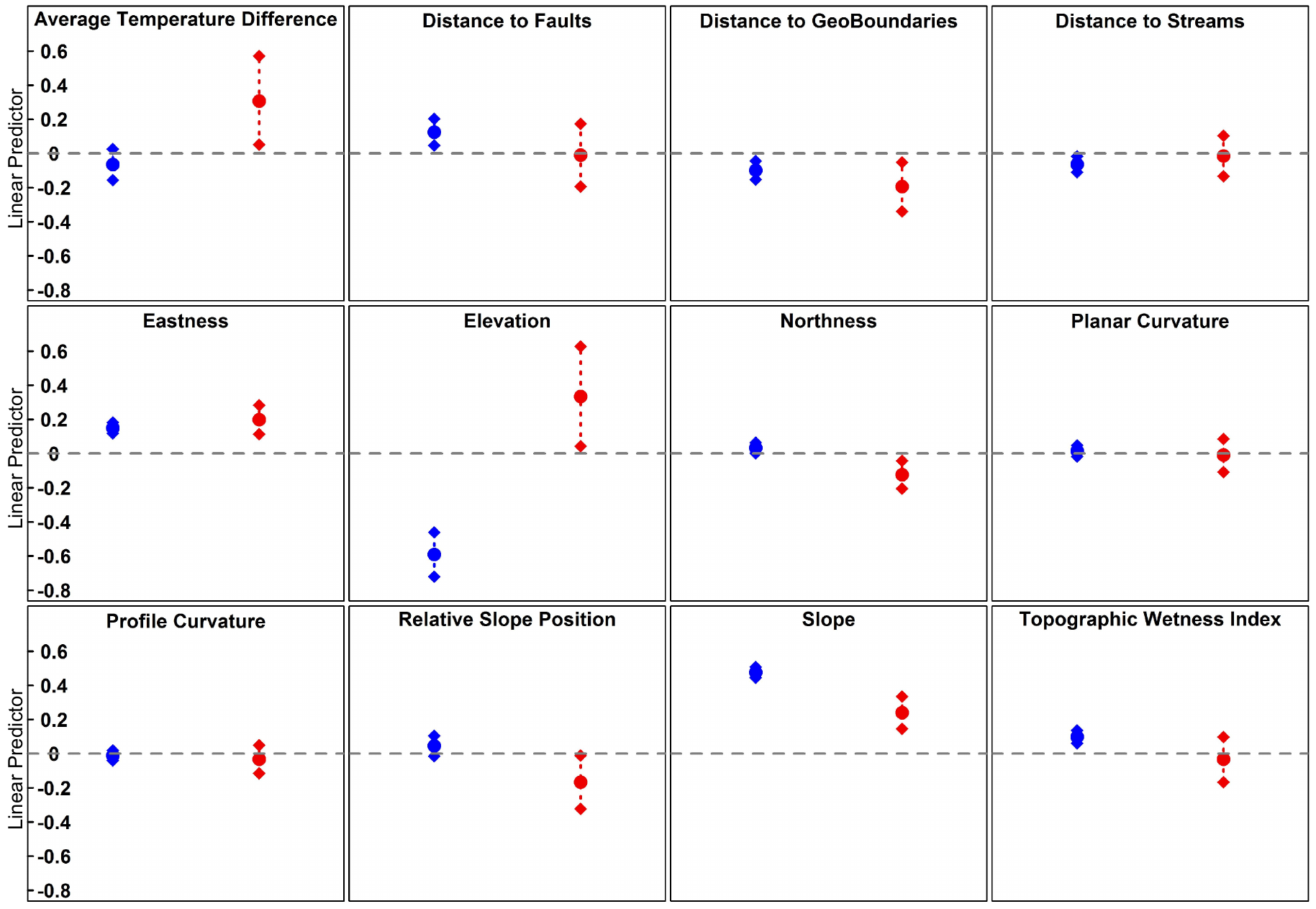}
\caption{Fixed effects for Lushan (blue) and Wenchuan (red) earthquakes. The y-axis represents the range of regression coefficients obtained for the covariates. For comparison, the covariates have all been rescaled with zero mean and unit variance. Negative coefficients decrease the landslide intensity whereas positive coefficients increase it. Coefficients lying on the zero line do not contribute to the model.}
\label{fig:Linearities}
\end{figure}

We report estimated linear effects in Figure \ref{fig:Linearities}. We found that: \textit{i}) \textit{Average Temperature Difference} is significant just the Wenchuan earthquake, contributing to an increase in the landslide intensity; \textit{ii}) \textit{Distance to Faults} is significant only for the Lushan earthquake and with a positive effect; \textit{iii}) \textit{Distance to GeoBoundaries} appears to be significant in both cases with a negative contribution to landslide counts; \textit{iv}) \textit{Distance to Streams} is significant only Lushan with a slightly negative contribution; \textit{v}) \textit{Eastness} is significant and positive in both cases; \textit{vi}) \textit{Elevation} presents a negative coefficient for Lushan and a positive one for Wenchuan; \textit{vii}) \textit{Northness} presents a positive coefficient for Lushan and a negative one for Wenchuan; \textit{viii}) \textit{Planar} and \textit{ix}) \textit{Profile Curvatures} are not significant; \textit{x}) \textit{Relative Slope Position} is not significant; \textit{xi}) \textit{Slope} is not only significant and positive, but also the strongest contributor to landslide intensity (recall that the covariates were all rescaled to make the coefficients directly comparable); and \textit{xii}) \textit{Topographic Wetness Index} is significant only for the Lushan earthquake with a positive effect. The interpretation of the predictors' role is provided in the Supplementary Material. 

\subsubsection{Combined Eastness and Northness Effects}
\label{sec:Aspect Effects}

\begin{figure}[t!]
\centering
\includegraphics[width=0.5\linewidth]{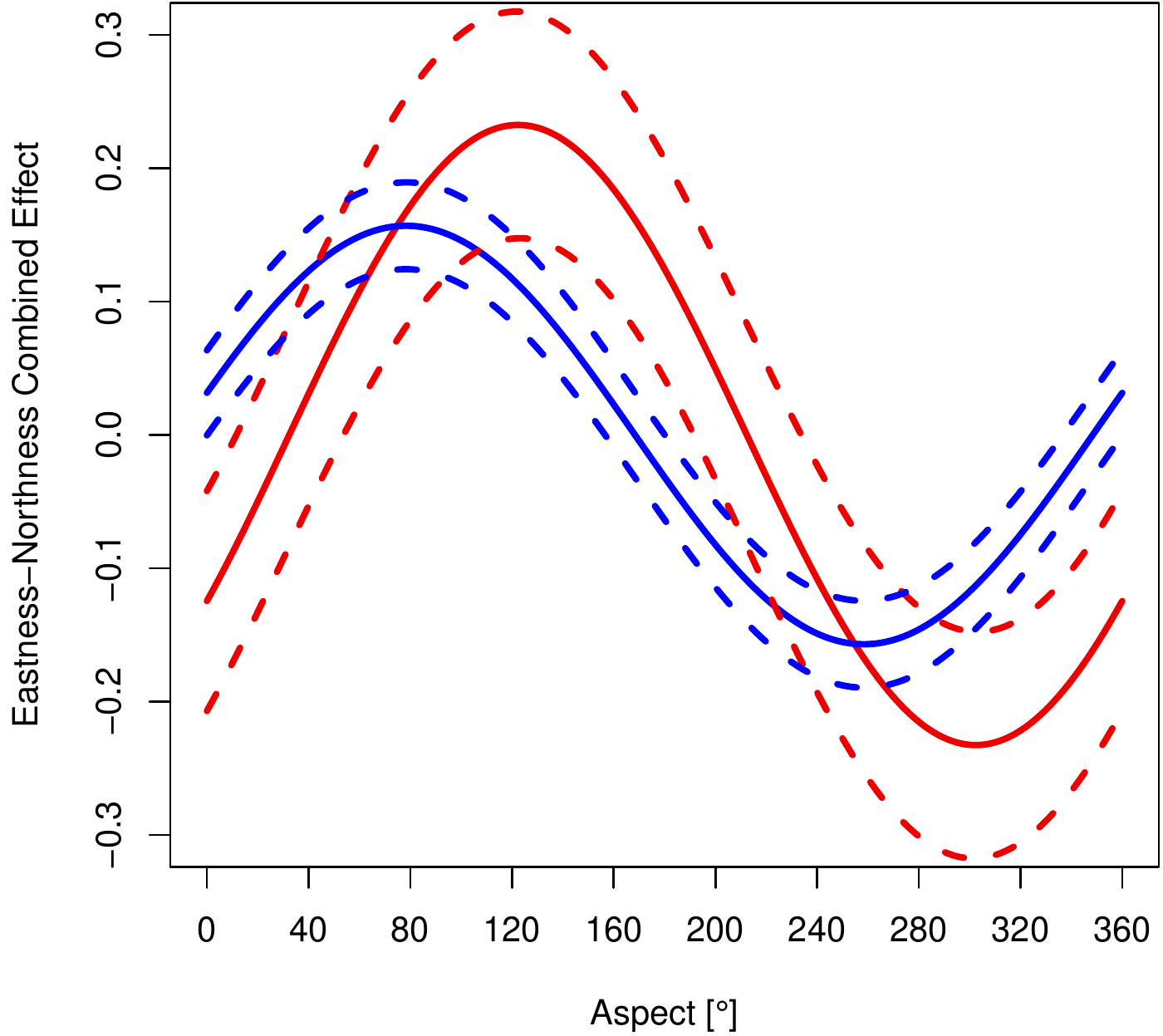}
\caption{Combined Eastness and Northness effect for the Wenchuan (red) and Lushan (blue) earthquakes. Solid lines represent the mean effect; dotted lines correspond to the 95\% CI.}
\label{fig:AspectEffect}
\end{figure}

The orientation of the slope with respect to the north (\textit{Aspect}) is cyclic by nature and is typically used either nonlinearly \citep{lombardo2015binary} as a categorical predictor or linearly via \textit{Eastness} and \textit{Northness} \citep[e.g.,][]{Steger2016}. In our model, we chose the second option (see Figure \ref{fig:Linearities}). Here, we present the overall effect of the Aspect as a sinusoidal function described by the combination of two fixed linear effects (i.e., we compute the the sum of $\sin(\mbox{\emph{Aspect}})$ and $\cos(\mbox{\emph{Aspect}})$ multiplied by the respective \emph{Eastness} and \emph{Northness} coefficients); thus, these effects can be interpreted in terms of the aspect itself. Figure \ref{fig:AspectEffect} shows the resulting curves for the two earthquakes (with their respective 95\% credible bands) to demonstrate that their effect is highly nonlinear and significant as a whole. Moreover, the Aspect effect for the Lushan and Wenchuan earthquakes are shown to be quite similar. An interpretation of the role of the Aspect is provided in the Supplementary Material.

\subsubsection{Random Effects}
\label{sec:Random Effects}

\begin{figure}[t!]
\centering
\includegraphics[width=\linewidth]{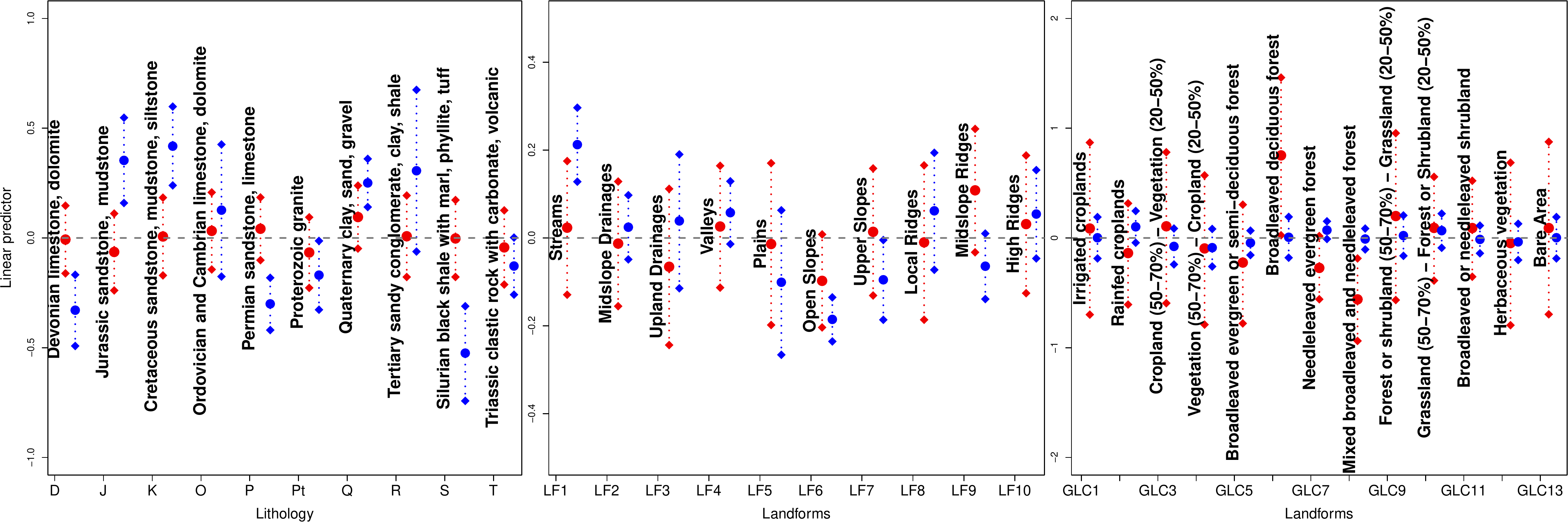}
\caption{Random effects for the Lushan (blue) and Wenchuan (red) earthquakes; acronyms for lithology are reported in accordance to \citet{Ding2015}.}
\label{fig:Categorical Effects}
\end{figure}

Figure \ref{fig:Categorical Effects} reports the estimated effects for each of the categorical covariates. Here, we report the effects for the two earthquakes separately. 

The only significant lithotype for the Wenchuan earthquake is \textit{Quaternary Clay, Sand, and Gravel}, which positively contributes to the landslide intensity. No significance can be seen among the landform classes, even if \textit{Open Slopes} (negative coefficient) and \textit{Midslope Ridges} (positive coefficient) miss significance by only a slight margin. Of the land cover classes, only \textit{Broadleaved Deciduous Forest} is significant and has a positive effect.

The Lushan earthquake has a larger number of significant effects, probably due to its larger number of observed landslides in our study area. \textit{Devonian Limestone, Dolomite}, \textit{Permian Sandstone, Limestone}, \textit{Proterozoic Granite} and \textit{Silurian Black Shale with Marl, Phyllite, Tuff} all decrease the estimated landslide counts. Conversely, \textit{Jurassic sandstone, mudstone}, \textit{Cretaceous Sandstone, Mudstone, Siltstone} and \textit{Quaternary Clay, Sand, Gravel} have positive effects in the model. \textit{Streams} and \textit{Open Slopes} are significant Landform classes with negative and positive effects, respectively. Surprisingly, no land cover class is shown to be significant, though \textit{Needleleaved Evergreen Forest} only slightly misses significance. Random effects are further discussed in the Supplementary Material.    

\section{Discussion}
\label{sec:Discussion}

The primary goal of this study was to test whether the LSE is able to capture the pattern of seismic shaking from the distribution of earthquake-induced landslides without having prior knowledge about the earthquake in the statistical model.
Our LSE-only model performed significantly better (see Table \ref{table2}) than the MI-only and no further improvements were achieved by using the LSE and MI together. 
These results support our initial hypothesis that the LSE can capture the signal of the trigger from multiple-occurrence regional landslide events \citep[MORLE,][]{crozier2005multiple}. 
This is graphically represented in Figure \ref{fig:LSE_VS_MI} where the MI and LSE maps (with very narrow credible intervals) present similar patterns for both earthquakes. This qualitative consideration was also confirmed via a more rigorous comparison. In Figure \ref{fig:ComparisonLSEandMI}, we report the relation between the MI (both in its original scale and its effect within the models) and the LSE. For the Lushan earthquake, the LSE behaved almost identically to the MI; more pronounced differences characterized the Wenchuan case. This may be attributed to the greater ability of the LSE to capture local effects such as topographic amplifications. 

To investigate this further, we compared the spatial patterns displayed by the MI and LSE belonging to the Lushan and Wenchuan earthquakes (see Figure \ref{fig:CrossSections}). Ground shaking may be amplified along crests due to topographic focusing of the incoming wavefield \citep{Imperatori2015} and, in response to this amplification, a greater number of slope failures may initiate \citep[e.g.,][]{Jafarzadeh2015}. However, in general, the ShakeMaps do not include this effect due to a lack of data from local ground stations \citep[e.g.,][]{Meunier2008}. Detailed investigation regarding the effects of surface topography on ground-shaking parameters will require numerical simulations \citep[e.g.,][]{Imperatori2015,Lee08,Lee09}, which have high computational costs. 

The results presented in Figure \ref{fig:CrossSections} indicate that our LSE may pick up part of the amplification at the crests, thus playing the role of the topographic focusing effect in the model. Conversely, the MI is too smooth to capture any amplification (particularly for the Lushan earthquake). In this regard, our approach can be considered an alternative method for ground-shaking estimation, especially, when reliable earthquake-induced landslide inventory are available but reliable earthquake source data are not, e.g., the 2015 Gorkha earthquake. \citet{Gallen2017} highlighted the relatively low quality of ShakeMap for Gorkha, while the associated landslides are comprehensively mapped \citep{Roback2017}.

The disadvantage of using the LSE is that it picks up any spatial residual signal in the data, thus making it difficult to recognize which properties are reflected in its pattern. For instance, the legacy of previous earthquakes can lower the rock-strength parameters in unfailed slopes \citep{Parker2015}. Therefore, we can assume that the shaking of the 2008 Wenchuan earthquake may have played a role in failures at sites affected by the 2013 Lushan earthquake. However, \citet{Tang2016} analyzed the post-earthquake landslide activity in the epicentral area of the Wenchuan earthquake and concluded that most of the landslides were not active after three years. They argued that the unfailed slopes weakened by the Wenchuan earthquake experienced sliding due to large amount of subsequent rainfall. Therefore, we assume that the contribution of the legacy effect from the Wenchuan earthquake on the LSE of the Lushan model is minor.  

Another novelty for the geomorphological literature is represented by the underlying distribution (Poisson) used to explain the landslide scenario, and the type of spatial model (log Gaussian Cox process) used to perform the geostatistical analysis. This distribution allowed us to jointly predict \emph{where} and \emph{how many} landslides are expected under similar triggering conditions (see Figure \ref{fig:Intensity maps}). Moreover, the \emph{intensity} maps we propose at the pixel level can be used to generate predictive maps for any mapping unit by summing all the intensity values for the pixels contained in a given polygon. This is a unique aggregative property of the intensity (using the Poisson distribution) that is not shared by its classical susceptibility counterpart (using the Bernoulli distribution). Through this aggregative property, we produced three more hierarchical intensity maps predicting landslide counts at slope, catchment, and administrative units.    

The overall performances appear to be outstanding, according to \citet{hosmer2000}, when we converted intensity to susceptibility both for the fitted models and the cross-validation procedure. Specifically, the model reached fitted AUC values of 0.889 and 0.943, and cross-validated AUC values equal to 0.850 and 0.928 at the pixel level, for the Lushan and Wenchuan events, respectively. Similarly, the intensity models also performed well both for the fitted models and CV procedures (see Table \ref{table3}). However, we note that when looking at the pixel results, the Poisson model did not produce equivalently good results with respect to the other mapping units. We attribute this to the disproportion between the few landslide counts and the numerous no-landslide pixels, which made the actual count dataset more similar to a binary one than to a typical Poisson case. Nevertheless, even at the pixel scale, our model predicted an overall global number of landslides equal to $8120$ for the Lushan earthquake and $987$ for the Wenchuan earthquake; the actual counts were $7868$ and $928$ landslides, respectively. This is a good indicator that the model is working well. However, in Poisson regression, the intensity is a single parameter that represents both the mean and variance of the original counts. For this reason, the model controls both the specific count value and how much it fluctuates stochastically in its neighborhood at the same time. Our interpretation of weak results at the pixel scale is due to the fact that the Poisson distribution ``struggles'' at very fine resolutions because the pixel intensity varies too much compared to the mean (a property known as ``overdispersion''). On the other hand, when we aggregated at a higher hierarchical mapping unit, the transition was smoother and gave rise to outstanding performances already at the slope unit level. 

Any statistical model requires a thorough interpretation of the covariate effects. We report the geomorphological interpretation in the Supplementary Material. 

\section{Conclusions}
\label{sec:Conclusions}

In this contribution, we show that the LSE is able to capture the spatial distribution of a trigger responsible for widespread landsliding (see Figure \ref{fig:LSE_VS_MI}), using two separate earthquake-triggered landslide inventories.
We demonstrate this by comparing the LSE to both the original seismic parameters and their final effects on the landslide intensity (see Figure \ref{fig:ComparisonLSEandMI}).
For the Lushan earthquake, the behavior of the LSE is almost exactly the same as the MI; for the Wenchuan earthquakes, slight differences can be seen.
When spatially predicting landslides, the shaking information is of crucial importance; however, this information may not always be available.
In this case, we suggest using the LSE to mimic the effect of unknown ground motion on landslide activation.
The LSE can also be applied to any predictive model where information about the main trigger is missing (e.g.,\ storms or snowmelt).

For earthquakes, there are cases where the shaking parameters are available but unreliable.
The USGS ShakeMaps publish a quality index \texttt{Empirical ShapeMap grade} that reports the reliability of shaking information.
For lower quality ShakeMaps, the LSE offers a valid alternative to including the earthquake effect into the model.
The LSE is completely independent from the seismic records and only requires a complete landslide inventory (which is the basic requirement for any predictive model).

Another strength of the method we propose is that an intensity model is much more informative than a susceptibility model.
In addition to predicting where new landslides may occur under similar triggering conditions, the intensity model estimates the potential landslide count.
This is not trivial in landslide susceptibility modeling, whereby a mapping unit with one landslide is treated the same way as a mapping unit with numerous landslides.
When a master planner examines a susceptibility map, the information conveyed in it does not reflect the number of potential activations. By contrast, an intensity map does. In turn, the probability that an object or person is hit during an earthquake event may be inferred from the landslide counts over space, which can be used as input in risk assessments. This further increases the value of intensity maps compared to susceptibility maps.

Institutions that deal with territorial management have a finite budget to spend on slope stabilization projects, and using intensity maps may help to prioritize investment in the highest risk areas.

From a methodological perspective, building susceptibility maps for different mapping units requires completely separate models, one for each mapping unit, with different
subjective assumptions on how to approximate the covariates at different scales.
However, the landslide intensity models we present are consistent across any spatial scale due to the aggregative properties of the Poisson distribution.

At this stage, the predicted landslide intensity includes the signal of a past earthquake, 
which will likely be a poor representation of a future one.
In the future, we plan to assess the spatial and temporal variability of the shaking patterns via several LSE simulations.
Each LSE will produce a different intensity map, allowing us to assess which part of the physiography exhibits the greatest landslide intensity, irrespective of earthquake directivity. Thus, the LSE will be used as an effective tool for predictive modeling in earthquake-induced landslide hazard and risk assessments.

\bibliographystyle{CUP}
\bibliography{landslides}

\includepdf[pages=-]{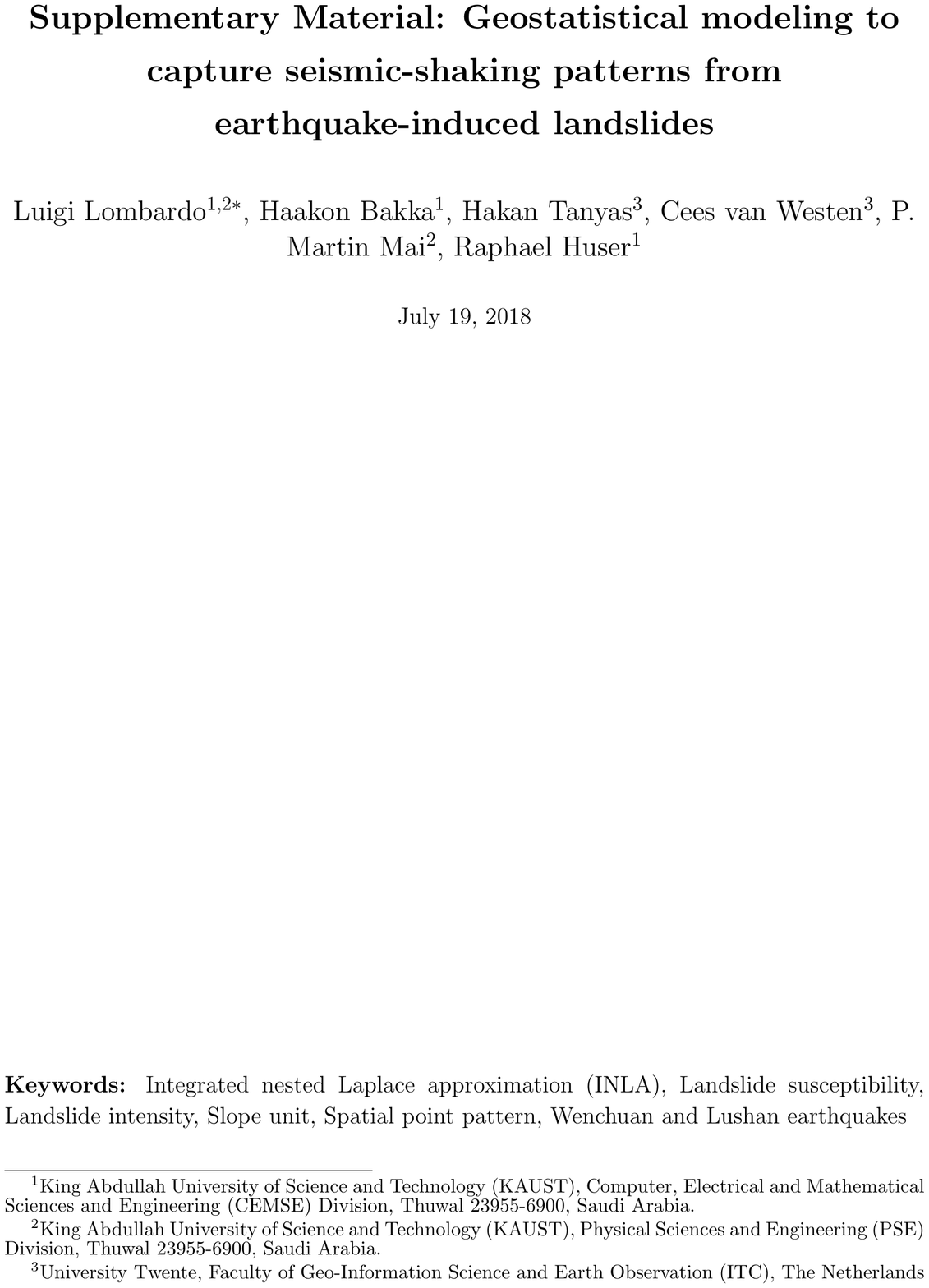}
\end{document}